# Improved self-energy correction method for accurate and efficient band structure calculation

Kan-Hao Xue,[1,2,*] Jun-Hui Yuan,[1,2] Leonardo R. C. Fonseca,[3,†] and Xiang-Shui Miao[1,2,‡]

[1]School of Optical and Electronics Information, Huazhong University of Science and Technology, Wuhan 430074, China
[2]Wuhan National Laboratory for Optoelectronics, Wuhan 430074, China
[3]Center for Semiconductor Components, University of Campinas, Campinas, São Paulo 13083-870, Brazil

ABSTRACT

The LDA-1/2 method for self-energy correction is a powerful tool for calculating accurate band structures of semiconductors, while keeping the computational load as low as standard LDA. Nevertheless, controversies remain regarding the arbitrariness of choice between (1/2)e and (1/4)e charge stripping from the atoms in group IV semiconductors, the incorrect direct band gap predicted for Ge, and inaccurate band structures for III-V semiconductors. Here we propose an improved method named shell-LDA-1/2 (shLDA-1/2 for short), which is based on a shell-like trimming function for the self-energy potential. With the new approach, we obtained accurate band structures for group IV, and for III-V and II-VI compound semiconductors. In particular, we reproduced the complete band structure of Ge in good agreement with experimental data. Moreover, we have defined clear rules for choosing when (1/2)e or (1/4)e charge ought to be stripped in covalent semiconductors, and for identifying materials for which shLDA-1/2 is expected to fail.

Density functional theory[1] (DFT) made practical by the Kohn-Sham ansatz[2] has been highly successful in first-principles calculations of materials properties, despite long-known and debated limitations[3,4] introduced by its local density or generalized gradient approximations[5] (LDA and GGA, respectively). Of particular relevance to the present work is the systematic underestimation of band gaps of semiconductors and insulators.[6] Indeed, since Koopmans' theorem is no longer valid under the approximations of the exchange-correlation potential, the Kohn-Sham eigenvalues, formally Lagrange multipliers for the constrained minimization of the Kohn-Sham total energy, cannot be simply interpreted as electronic energy levels. Solutions to this problem involve improving the exchange energy by introducing a certain amount of accurate Hartree-Hock exchange (the hybrid functional scheme[7]) or by taking a different path with the adoption of the quasi-particle approach within the GW approximation,[8] both methods at a computational load substantially higher than LDA. Recently, a new self-energy correction method named LDA-1/2 was proposed by Ferreira and coworkers, which yields accurate band structures for many compounds, is as computationally efficient as LDA, and fully *ab-initio*.[9,10] LDA-1/2 is similar to SIC[11] as both methods attempt to solve the band gap problem by correcting the self-energy. However, in addition to its computational efficiency, LDA-1/2 is much simpler to implement. The recently proposed $\Delta$-sol[12] is another efficient technique for calculating accurate band gaps. However, it does not correct the full band structure, only the fundamental band gap, and is not as straightforward as LDA-1/2 since it requires several total energy calculations.

LDA-1/2 stems from Slater's proposal of a transition state in his X$\alpha$ method to reduce the band gap inaccuracy from second to third order by introducing a half-electron/half-hole occupation, the so-called transition state.[13] Ferreira *et al.* extended this method to modern DFT and particularly to solid-state, under the assumption that the excited electron in the conduction band of a semiconductor usually occupies Bloch-like states with nearly-null self-energy, while the hole left in the valence band is localized with a finite self-energy.[9] In order to correct the self-energy of the localized hole in a solid, they introduced a self-energy potential calculated for the atoms associated with the top of the valence band and modified the corresponding pseudopotentials, which are then employed in solid state calculations. According to the original technique, for ionic bonds the self-energy potential is the difference between a neutral atom and its ion with 1/2 electron stripped for the anion, following Slater's half-occupation technique for isolated atoms. When the hole is shared by two atoms in covalent bonds, 1/4 electron is stripped from each ion to avoid correcting twice. In this work the name LDA-1/2 is reserved for the technique in general, while we denote each particular implementation of LDA-1/2 by two minus signs followed by the amount of charge stripped from each of the atoms forming their chemical bond. For example, the ionic and covalent bonds mentioned above are here called LDA-0-1/2 and LDA-1/4-1/4, respectively, where "0" and "1/2" are the amount of elementary charge stripped from the cation (A) and the anion (B) in the ionic semiconductor AB, while "1/4" is the amount of elementary charge stripped from the two atoms forming the covalent bond, either between identical (such as in Si-Si) or distinct species (such as in Ga-As).

Despite its successful application to numerous compounds, the original LDA-1/2 (or GGA-1/2, which is implemented the same) method still suffers from ambiguities, especially regarding covalent semiconductors. For example, to properly reproduce experimental band gaps in simple covalent semiconductors Ferreira *et al*. suggested stripping (1/4)e from the 3s and 3p states of diamond carbon (LDA-[(1/4)s+(1/4)p]-[(1/4)s+(1/4)p]), while for diamond silicon LDA-(1/4)p-(1/4)p should be used instead.[9] Our calculations indicate that using LDA-(1/2)p-(1/2)p (LDA-1/2-1/2 for short) or LDA-[(1/4)s+(1/4)p]-[(1/4)s+(1/4)p] for carbon yields very similar band gaps. Hence, carbon and silicon require different treatments ((1/2)e or (1/4)e stripped from every C or Si atom, respectively), even if their structures are highly isomorphic. Moreover, both LDA-1/4-1/4 and LDA-1/2-1/2 incorrectly predict germanium to be a $\Gamma$-to-$\Gamma$ direct band gap semiconductor.[14] Last but not least, there are cases when LDA-1/2 works very well for a particular compound but fails for another one with similar properties. For instance, LDA-1/2 predicts excellent band gap for ZnO, but severely underestimates (despite improving over LDA) the band gap for $Cu_2O$. Due to its low computational cost, it is natural not to expect LDA-1/2 to be so comprehensive as to fit all semiconductors. However, a deeper understanding of the method should offer more satisfactory rules for when and why LDA-1/2 is expected to fail. In particular, the apparently arbitrary choice of orbital from which to strip charge and its amount hinders the widespread use of the method. In this Letter we propose a modification of the original method to improve its accuracy, we explain unambiguously the correct choice of LDA-1/2-1/2 and LDA-1/4-1/4 for covalent semiconductors, and provide a rule for judging the applicability of LDA-1/2 in general.

As mentioned before, the original LDA-1/2 formulation should work best when the hole is localized in small regions of real space, while the excited electron is delocalized. Without accounting for the hole self-energy, the energy of the Kohn-Sham highest occupied state is thus too high, leading to the underestimation of the band gap. Although the calculation of a rectifying hole self-energy in k-space in rather complex, according to Ferreira *et al.* it is much simpler to model in real space, by associating an atomic-derived self-energy potential to those regions where the hole is localized. Under these conditions, LDA-1/2 works as an accurate and computationally efficient scheme for correcting LDA band gaps. A broad array of examples can be found elsewhere.[10]

In the original LDA-1/2 the hole self-energy is obtained from the difference of atomic potentials between an isolated neutral atom and its ion with 1/2 electron stripped, which is then added to the pseudopotential of the corresponding atom (i.e., the anion) in periodic solid state calculations. To keep the self-energy correction local, a trimmed self-energy potential $V_s$ is used instead, which in the original proposal is written as

$$V_s = V_s^0 \Theta(r) = \begin{cases} V_s^0 \left[1 - \left(\frac{r}{r_{\text{cut}}}\right)^8\right]^3 & , r \leq r_{\text{cut}} \\ 0 & , r > r_{\text{cut}} \end{cases} \quad (1)$$

where $V_s^0$ is the unscreened atomic self-energy potential,[9] $\Theta$ is a trimming function, and $r_{\text{cut}}$[15] is obtained variationally upon maximizing the band gap of the corresponding bulk oxide. Since the pseudopotential is *r*-dependent, a spherical trimming function as the one defined in Eq. 1 seems natural. However, in general the hole may be found not centered at the anions but in the region between anions, mostly along the bonds. Therefore, as a more generic scheme for correcting the self-energy only where the hole is concentrated, we propose to trim the self-energy potential employing a spherical shell whose inner radius $r_{\text{in}}$ ranges from zero (where the new scheme is very similar to the original LDA-1/2) to an outer cutoff radius $r_{\text{out}}$. Both $r_{\text{in}}$ and $r_{\text{out}}$ radii are obtained variationally. We call this technique shell-LDA-1/2 (shLDA-1/2).

To show how the shell cutoff function compares with the original spherical cutoff, we calculated the band structure and the partial charge density of several benchmark semiconductors. DFT calculations were carried out using the plane-wave based Vienna *Ab Initio* Simulation Package (VASP) code,[16,17] with a 600 eV plane-wave energy cutoff. Standard LDA within the form of Ceperly-Alder was used as the exchange-correlation functional,[18] while for the GGA calculations we used the PBEsol functional.[19] The valence electrons were approximated by projector augmented-wave pseudopotentials.[20,21] In case of GGA calculations we used the PBEsol functional. All structures were fully relaxed until the stress in all directions were less than 100 MPa and all Hellmann-Feynman forces were less than 0.01 eV/Å. Self-energy corrections to the band structures were carried out utilizing the structures optimized with LDA.

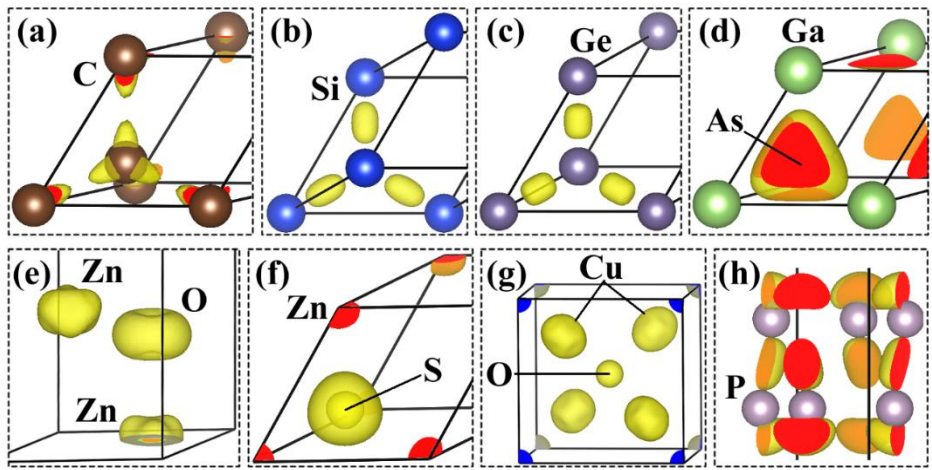

Figure 1. Hole charge density in (a) diamond carbon; (b) silicon; (c) germanium; (d) GaAs; (e) wurtzite ZnO; (f) zinc blende ZnS; (g) $Cu_2O$; (h) single-layer black phosphorus. See text for details.

For all materials analyzed, the first step consisted of a detailed partial charge density analysis to identify the region of highest hole concentration in real space. We start with the group IV semiconductors C, Si and Ge, all in the diamond structure. Figure 1 displays the calculated partial charge density of the topmost valence band near the k-point where the hole resides in the reciprocal space (Γ for the three materials). Figure 1(a) shows that in C the hole is located at four symmetric regions closely surrounding the C-atom, while in Si and Ge it is located near the bond centers (Figs.

1(b) and 1(c)). These results suggest that a shell-like trimming of the self-energy potential can be more effective than a spherical trimming in the cases of Si and Ge since it avoids an undesired correction of the near-core region where the probability of finding the hole is low. On the other hand, for C the use of a shell-like trimming should result in a zero inner radius after optimization since the probability of finding the hole near the atom is high. We have adopted the following formula for the shell trimming function

$$V_s = V_s^0 \Theta(r) = \begin{cases} V_s^0 \left[1 - \left(\frac{r}{r_{out}}\right)^m\right]^3 \frac{1+\tanh[n(r-r_{in})]}{2} &, r \leq r_{out} \\ 0 &, r > r_{out} \end{cases} \quad (2)$$

where $m$ and $n$ should be sufficiently large to ensure a sharp trim in both the shell inner and the outer regions. In this work we have used $m = n = 50$. Larger values were also tried with little effect on the results. Notice that the slope of the *tanh* function is usually larger than the slope of polynomials, therefore the function above provides a sharper trim for the inner than for the outer region, which is reasonable since the potential and electron density vary faster near the atomic core.

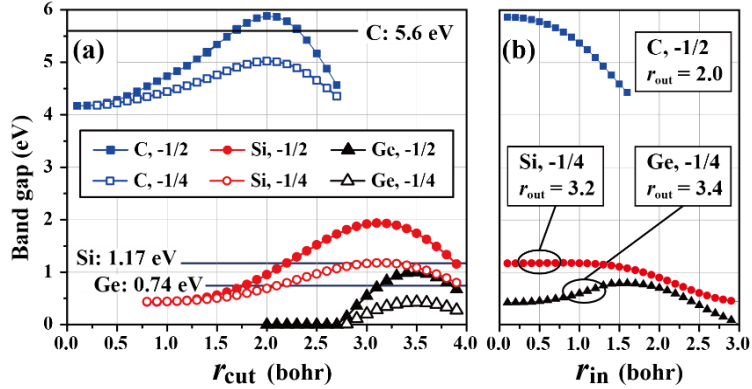

Figure 2. Calculated band gaps versus (a) the cutoff radii $r_{cut}$ in LDA-1/2 (Eq. 1) and (b) the inner cutoff radii $r_{in}$ in shLDA-1/2 (Eq. 2) for group IV semiconductors. The outer cutoff radii $r_{out}$ optimization of shLDA-1/2 using Eq. 2 (not shown) gives almost identical results as in (a) and assumed a spherical trimming function ($r_{in} = 0$), while the $r_{in}$ were varied in (b) using the optimized shLDA-1/2 $r_{out}$.

In Fig. 2 we show the cutoff radii optimization for C, Si and Ge using LDA-1/2 and shLDA-1/2. Figure 2(a) compares the LDA-1/4-1/4 and LDA-1/2-1/2 band gaps as the cutoff radii are varied assuming a spherical trimming function (similar to shLDA-1/4-1/4 or shLDA-1/2-1/2 with $r_{in}$ = 0). In all cases, the optimal $r_{cut}$ is the same for LDA-1/4-1/4 and LDA-1/2-1/2. LDA-1/2 requires the subtraction of half electron from each region where the hole is found. As shown in Fig. 1(a), in bulk C the hole density is mostly located in the vicinity of C atoms, extending only about 1/3 of the C-C bond length. Hence, subtracting (1/2)e applies for diamond C while subtracting (1/4)e does not as the hole cloud is not shared by two C atoms. Indeed, the band gap for C obtained with LDA-1/2-1/2 (5.88 eV) is closer to experiment (5.6 eV) than the band gap obtained with LDA-1/4-1/4 (5.02 eV). The optimal $r_{cut}$ in this case is around 2 bohr, approximately 2/3 of the C-C bond length (2.9 bohr). For Si and Ge the spatial distribution of the hole density is quite different, being concentrated mostly around the bond centers. Consequently, by symmetry the self-energy in these regions is corrected twice, making the stripping of (1/4)e from all atoms the valid choice. LDA-1/4-1/4 together with a larger outer cutoff radius of 3.2 (3.4) bohr for Si (Ge) guarantees the double counting of bond centers.

The calculated band gap for Si confirms the arguments above, with the LDA-1/4-1/4 band gap matching the experimental value, 1.17 eV, while the LDA-1/2-1/2 band gap, 1.93 eV, is overestimated. For Ge, however, the LDA-1/4-1/4 band gap is much smaller than the experimental value, despite the improvement over plain LDA which yields a zero band gap.[6] This difficulty is caused by the correction of the self-energy near the atoms where the hole density is low (Fig. 1c), suggesting that shLDA-1/4-1/4 might be more accurate. As shown in Fig. 2(b), allowing $r_{in}$ to increase from zero keeps the band gap of Si almost unchanged, while in C and Ge the band gap decrease and increase, respectively. shLDA-1/4-1/4 also requires that $r_{in}$ and $r_{out}$ are such that the band gap is maximum. Thus in C and Si shLDA-1/2-1/2 and shLDA-1/4-1/4 reduce to standard LDA-1/2-1/2 and LDA-1/4-1/4, without the need of an inner cutoff radius. However, for Ge the band gap maximum, 0.79 eV, occurs for $r_{in}$ = 1.6 bohr, showing that Ge, contrary to Si, is sensitive to the unnecessary correction of the self-energy near the atoms caused by a spherically trimmed self-energy potential.

To verify this point further, we plot the band diagrams of Ge in Figs. 3(a) and 3(b), employing LDA-1/4-1/4 and shLDA-1/4-1/4. While LDA-1/4-1/4 predicts Ge to be a direct Γ-to-Γ band gap semiconductor, shLDA-1/4-1/4 predicts Ge to be a Γ-to-L indirect gap semiconductor. Moreover, both direct and indirect band gaps for Ge obtained with shLDA-1/4-1/4, 0.88 eV and 0.79 eV, are in better agreement with experimental data (0.89 eV and 0.74 eV), than obtained with LDA-1/4-1/4, 0.43 eV and 0.62 eV, respectively.

GaAs is another covalent semiconductor which presents difficulties for the spherically trimmed LDA-1/2. The hole location shown in Fig. 1(d) suggests stripping (1/2)e from As only, but the strongly non-spherical charge cloud reflects the influence of Ga. However, such influence is not enough to justify a self-energy correction for Ga due to the negligible hole density close to it. Thus, it is LDA-0-1/2 rather than LDA-1/4-1/4 that must be used in GaAs. However, in the context of shLDA-1/2, the hole presence along the Ga-As bonds, though more concentrated around As, suggests that shLDA-1/4-1/4 is the proper way for correcting the self-energy in this material. Finally, the optimized inner and outer cutoff radii for Ga are 2.2 bohr and 3.9 bohr, respectively, forming a narrow shell with a considerable uncorrected region centered around Ga and supporting the conclusion that 1/4 electron removal is only justifiable in the context of shLDA-1/2 rather than LDA-1/2. As shown in Figs. 3(c) and 3(d), not only for the direct Γ-to-Γ gap, but also for the indirect band gaps along the L and X directions, do we see great improvement brought by shLDA-1/4-1/4 over LDA-0-1/2. In particular, the calculated band gap along X is 2.41 eV with LDA-0-1/2, which is more than 0.4 eV higher than the experimental value, while the shLDA-1/4-1/4 error is merely 0.12 eV.

Next, we calculate the band gaps of several important semiconductors (see Table I) using shLDA-1/2 to demonstrate the generality of the arguments used previously. Table I shows that the overall improvement of shLDA-1/2 over LDA-1/2 is considerable. The special failure case is GaSb, where shLDA-1/4-1/4 yields similar indirect Γ-to-L and direct Γ-to-Γ band gaps, both 0.14 eV larger than the experimental direct band gap. However, we have tested LDA against GGA for Sb- and Te-based compounds and found that GGA yields considerably better band structures. Thus, we re-calculated GaSb using GGA-1/4-1/4 and shGGA-1/4-1/4 to correct the GGA band structure and found with the latter a direct band gap of 0.82 eV, extremely close to the experimental value of 0.81 eV. In addition, the L-to-Γ energy difference for the conduction band obtained with shGGA-1/4-1/4 is 0.084 eV, also very close to the experimental value of 0.089 eV.

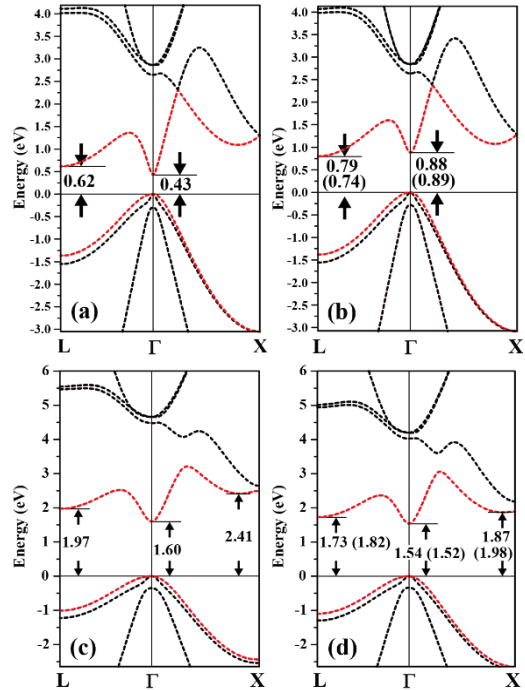

Figure 3. Band diagrams for (a) Ge calculated with LDA-1/4-1/4; (b) Ge calculated with shLDA-1/4-1/4; (c) GaAs calculated with LDA-0-1/2; (b) GaAs calculated with shLDA-1/4-1/4. The experimental band gap values are included in parentheses for comparison.

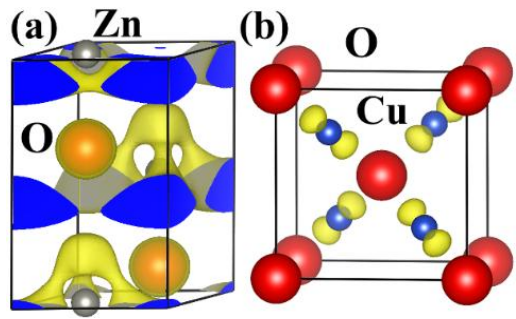

Figure 4. Charge density for the excited conduction band electron in (a) ZnO; (b) $Cu_2O$.

For the II-VI semiconductors considered in Table I, the band gaps calculated with shLDA-1/2 show considerable improvement over LDA-1/2. For all cases the hole is localized around the anions except for ZnO where, as already indicated by Ferreira et al.,[9,10] the hole is localized both near O and Zn sites as shown in Fig. 1(e). In contrast, in ZnS the hole is most likely found in a shell-region surrounding the sulfur atoms only. Therefore, to properly correct the hole self-energy in ZnO,

shLDA-1/2-1/2 (half electron stripped from Zn 3d and from O 2p) should be employed. There is no overcorrection of the hole self-energy in this case since both Zn and O equally concentrate the hole charge, without charge overlap. To further verify the apparent contradiction of correcting a cation ($Zn^{2+}$) self-energy, in Fig. 4(a) we plot the most likely spatial location for the conduction band electron in ZnO. Clearly, the conduction band electron surrounds the Zn atoms, but at a larger radial distance from the core, leaving the near core region for the hole, which requires self-energy correction. For ZnS, ZnSe, and ZnTe, on the other hand, the hole is localized only near the anions, thus making the self-energy correction of the anion sufficient. However, with optimized $r_{in}$ as large as 1.5 bohr as in the case of ZnTe, the shell trimming function provides improved band structures for the II-VI semiconductors.

shLDA-1/2, similarly to the original LDA-1/2, only involves generating modified pseudopotentials for band structure calculations. However, since the inner and outer cutoff radii vary depending on the chemical environment, we emphasize that for distinct materials the set of cutoff radii can be different for the same element. For example, $r_{in}$ ($r_{out}$) are 1.3 (3.4) bohr and 2.2 (3.9) bohr for Ga in GaP and GaAs, respectively. Hence, each material requires a separate optimization of the cutoff radii and the modified pseudopotentials are not transferable in general. Table I. Band gap values for typical group IV, III-V, II-VI, and other relevant semiconductors. *i* and *d* stand for indirect and direct band gaps, respectively.

| Material | Band gap (eV) | | | | |
|---|---|---|---|---|---|
| | LDA-0-1/2 | LDA-1/4-1/4 | shLDA-0-1/2 | shLDA-1/4-1/4 | Experimental |
| C (Diamond) | 5.88* (*i*) | - | 5.88† (*i*) | - | 5.6 (*i*) |
| | Hole localized mostly around each C atom. $r_{in} = 0$, $r_{out} = 2.0$ bohr. Units are the same below. | | | | |
| Si | - | 1.17 (*i*) | - | 1.18 (*i*) | 1.17 (*i*) |
| | Hole along the Si-Si bonds. $r_{in} = 0.8$, $r_{out} = 3.1$. | | | | |
| Ge | - | 0.43 (*d*) | - | 0.79 (*i*) <br> 0.88 (*d*) | 0.74 (*i*)[22] <br> 0.89 (*d*)[22] |
| | Hole along the Ge-Ge bonds. $r_{in} = 1.6$, $r_{out} = 3.5$. | | | | |
| GaP | 2.67 (*i*) <br> 3.00 (*d*) | - | - | 2.23 (*i*) <br> 2.63 (*d*) | 2.34 (*i*)[22] <br> 2.87 (*d*)[23] |
| | Hole along Ga-P bonds, closer to P. $r_{in}$ = 1.3, $r_{out}$ = 3.4 for Ga; $r_{in}$ = 0.7, $r_{out}$ = 2.9 for P. | | | | |
| GaAs | 1.60 (*d*) | - | - | 1.54 (*d*) | 1.52 (*d*)[22] |
| | Hole along Ga-As bonds, closer to As. $r_{in}$ = 2.2, $r_{out}$ = 3.9 for Ga; $r_{in}$ = 1.5, $r_{out}$ = 3.2 for As. | | | | |
| GaSb | 0.85 (*d*) | - | - | 0.82‡ (*d*) | 0.81 (*d*)[22] |
| | Hole along Ga-Sb bonds, closer to Sb. $r_{in}$ = 2.2, $r_{out}$ = 4.0 for Ga; $r_{in}$ = 1.9, $r_{out}$ = 3.6 for P. | | | | |
| AlAs | 2.82 (*i*) | - | - | 2.20 (*i*) | 2.23 (*i*)[22] |
| | Hole along Al-As bonds, closer to As. $r_{in}$ = 0.4, $r_{out}$ = 3.3 for Al; $r_{in}$ = 0.9, $r_{out}$ = 3.2 for As. | | | | |
| AlSb | 2.08 (*i*) | - | - | 1.65 (*i*) | 1.69 (*i*)[22] |
| | Hole along Al-Sb bonds, closer to Sb. $r_{in}$ = 1.8, $r_{out}$ = 3.8 for Al; $r_{in}$ = 1.1, $r_{out}$ = 3.4 for Sb. | | | | |
| InP | 1.79 (*d*) | - | - | 1.54 (*d*) | 1.42 (*d*)[22] |
| | Hole along In-P bonds, closer to P. $r_{in}$ = 2.5, $r_{out}$ = 4.1 for In; $r_{in}$ = 1.5, $r_{out}$ = 3.2 for P. | | | | |
| InAs | 0.78 (*d*) | - | - | 0.61 (*d*) | 0.41 (*d*)[22] |
| | Hole along In-As bonds, closer to As. $r_{in}$ = 2.5, $r_{out}$ = 4.4 for In; $r_{in}$ = 1.5, $r_{out}$ = 3.3 for As. | | | | |
| ZnO | 3.33* (*d*) | - | 3.36† (*d*) | - | 3.40 (*d*)[24] |
| | Hole localized on Zn and O atoms, no overlap. $r_{in}$ = 0, $r_{out}$ = 1.4 for Zn; $r_{in}$ = 0.5, $r_{out}$ = 2.2 for O. | | | | |
| ZnS | 3.52 (*d*) | - | 3.79 (*d*) | - | 3.84 (*d*)[22] |
| | Hole localized on S atoms. $r_{in}$ = 1.1, $r_{out}$ = 2.8. | | | | |
| ZnSe | 2.53 (*d*) | - | 2.79 (*d*) | - | 2.83 (*d*)[22] |
| | Hole localized on Se atoms. $r_{in}$ = 1.2, $r_{out}$ = 3.0. | | | | |
| ZnTe | 2.24 (*d*) | - | 2.50 (*d*) <br> 2.47†† (*d*) | - | 2.39 (*d*)[22] |
| | Hole localized on Te atoms. $r_{in}$ = 1.5, $r_{out}$ = 3.3. | | | | |
| Cu$_2$O | 1.00 (*d*) | - | 1.00 (*d*) | - | 2.17 (*d*) |
| | Hole localized on Cu atoms. $r_{in}$ = 0, $r_{out}$ = 2.3. | | | | |
| Li$_2$O$_2$ | 2.89 | - | 3.02 | | #4.91[25] |
| | Hole located on O atoms. $r_{in}$ = 0.8, $r_{out}$ = 1.9. | | | | |
| black P (single layer) | 1.24 (*d*) | 0.92 (*d*) | 1.25 (*d*) | - | #1.6 (*d*)[26] |
| | Hole localized both on the vacuum sides of the P monolayer and along the P-P bonds. $r_{in}$ = 0.7, $r_{out}$ = 2.8. | | | | |

\* LDA-1/2-1/2    † shLDA-1/2-1/2    ‡ shGGA-1/4-1/4    †† shGGA-1/2

# We could not find any reliable experimental band gap; the value here is from GW calculation

Despite the improvements brought by shLDA-1/2 to the band structures of important families of semiconductors, we now discuss some particular cases when both LDA-1/2 and shLDA-1/2 are expected to fail. These especial occurrences share the same general shortcomings, which are: (1) one cannot apply the self-energy correction without affecting states that do not suffer from self-interaction, and (2) the same atom requires different self-energy correction approaches due to diverse hole charge density properties in real space. $Cu_2O$ is a good example for case (1), where the electron in the bottom of the conduction band and the hole in the top of the valence band share the same region of space in a way that cannot be separated by a spherically symmetric trimming function. As shown in Figs. 1(g) and 4(b), the hole and the excited electron occupy the main diagonal of the $Cu_2O$ unit cell. While the electron and hole in fact do not overlap, any spherically symmetric trimming function cannot capture the hole without strongly disturbing the conduction band electron. Consequently, even shLDA-1/2-1/2 predicts a band gap much lower than experimental value. We have encountered this situation also for $O^-$ in $Li_2O_2$ where the O-Li $p_{xy}$ orbitals contribute most to the states near the top of the valence band, thus requiring self-energy correction, while the O-O $p_z$ orbitals form a narrow empty band, which does not ask for self-energy correction. Since these orbitals are associated with the same O atom, the self-energy correction applied to O-Li $p_{xy}$ orbitals cannot be decoupled from the O-O $p_z$ orbitals, lowering the energy of the valence band and of the O-O empty band. Black phosphorus illustrates case (2), when the correction of the self-energy requires $(1/2)e$ and $(1/4)e$ stripping from the same atom. Figure 1(h) shows the hole location, where for each P atom there is a high hole density distribution on the vacuum sides of the atomic plane. In the vacuum region the non-overlap of hole charge requires stripping $(1/2)e$ and a small $r_{out}$. Yet, a high hole density distribution is also observed along the P-P bonds, requiring $(1/4)e$ stripping and a relative large $r_{out}$. The different self-energy correction requirements of the two regions cannot be reconciled with a set of unique $r_{in}$ and $r_{out}$. Hence, the LDA-1/2 and shLDA-1/2 band gaps for black P are severely underestimated.

Finally we discuss a possible extension of shLDA-1/2 to allow for stripping different charge values from the ions, other than $(1/2)e$ or $(1/4)e$. This is applicable only for those cases when the hole is shared by the two ions forming the bond. Indeed, the choices of shLDA-0-1/2 and shLDA-1/4-1/4 (but not shLDA-1/2-1/2 when the hole is not shared) are two extreme cases of a general shLDA-x-y where x+y=1/2, just as purely ionic and covalent bonds are idealized chemical pictures. For example, GaP bonding is more ionic than GaAs, suggesting that the optimal charge stripping may not be (1/4)e from both atoms, but more charge stripped from P than from Ga. We have tested shLDA-x-y and found an even better agreement with experimental data than with the fixed charge stripping scheme. For example, in the case of GaP the optimal configuration is shLDA-0.19-0.31 which exactly recovers the experiment band gap 2.34 eV, while for shLDA-1/4-1/4 the disagreement is 0.11 eV. Despite its higher accuracy, the generalized fractional charge stripping technique involves fitting to the experimental band gap, contrary to the fixed charge stripping which has the great benefit of being fully *ab initio*.

In conclusion, employing hole localization analysis we explained the previous arbitrariness in LDA-1/2 with respect to the amount of charge to strip from the atoms forming a covalent bond, if (1/2)e or (1/4)e. Furthermore, we proposed a new LDA-1/2-like method, named shell-LDA-1/2 (shLDA-1/2 for short), which utilizes a shell-like trimming function to limit the extension of the atomic self-energy potential, thus avoiding overlap with the neighbors, and the undesired correction of the self-energy near the atomic cores when the hole is only concentrated in the region between atoms. The new technique shows great improvement over the original LDA-1/2 results based on a sphere-like trimming function. shLDA-1/2 correctly recovers the indirect band gap of germanium, predicts with great accuracy the band gaps of typical III-V and II-VI semiconductors, within 0.2 eV from experimental values, and gives quantitative support to the recipe for determining the amount of charge to strip from each atom forming covalent bonds. A generic rule for correctly applying shLDA-1/2 is prescribed, and representative examples illustrating its failure are discussed, indicating clearly the circumstances leading to its failure. sh-LDA-1/2 is trivially implemented in a few steps: (1) relax the lattice constants and atomic positions using LDA; (2) determine the amount of charge to strip from one or more species from the partial charge density analysis; (3) generate the self-energy potentials with various inner and outer cutoff radii and sum them to the pseudopotentials; (4) run self-consistent small bulk calculations with these modified pseudopotentials and select the optimum set of cutoff radii which maximize the band gap. The selected modified pseudopotential yields the correct bulk band structure and can be employed in other electronic structure calculations for the same material involving more complex structures, such as surfaces and interfaces. Our results indicate that shLDA-1/2 is a powerful tool for calculating accurate band structures of semiconductors, while demanding low computational load, comparable to standard LDA.

**Acknowledgement**

This work was supported by the MOST of China under Grant No. 2016YFA0203800, the Natural Science Foundation of Hubei Province under Grant No. 2016CFB223, and the Fundamental Research Funds for the Central Universities of China under Grant No. HUST:2016YXMS212. L.R.C. Fonseca thanks the Brazilian agency CNPq for financial support. The authors thank R. Ramprasad for helpful discussions and suggestions.

*Email address: xkh@hust.edu.cn

†Email address: fonsecalrc2@gmail.com

‡Email address: miaoxs@hust.edu.cn

---